# A Review of Data Mining in Personalized Education: Current Trends and Future Prospects

**Zhang Xiong**[a, b]**, Haoxuan Li**[a]**, Zhuang Liu**[a]**, Zhuofan Chen**[a]**, Hao Zhou**[a]**, Wenge Rong**[a]**, Yuanxin Ouyang**[a]

[a] School of Computer Science and Engineering, Beihang University, Beijing 100191, China

[b] School of Information Technology and Management, University of International Business and Economics, Beijing 100029, China

**Abstract** Personalized education, tailored to individual student needs, leverages educational technology and artificial intelligence (AI) in the digital age to enhance learning effectiveness. The integration of AI in educational platforms provides insights into academic performance, learning preferences, and behaviors, optimizing the personal learning process. Driven by data mining techniques, it not only benefits students but also provides educators and institutions with tools to craft customized learning experiences. To offer a comprehensive review of recent advancements in personalized educational data mining, this paper focuses on four primary scenarios: educational recommendation, cognitive diagnosis, knowledge tracing, and learning analysis. This paper presents a structured taxonomy for each area, compiles commonly used datasets, and identifies future research directions, emphasizing the role of data mining in enhancing personalized education and paving the way for future exploration and innovation.

**Keywords** personalized education, data mining, educational recommendation (ER), cognitive diagnosis, knowledge tracing, learning analysis

## 1 Introduction

Personalized education, aligning with contemporary educational trends, customizes learning to each student's unique needs, preferences, and capabilities. The rise of personalized education, intertwined with the digital age, leverages educational technology as a key component of learning environments. This evolution is supported by the rapid development and widespread utilization of artificial intelligence (AI), significantly enhancing the efficacy of personalized education. Meanwhile, the collection of vast amounts of data on educational platforms, combined with advanced techniques, particularly data mining, has opened new avenues for understanding and optimizing various aspects of students' learning process (e.g., individual learning trajectories, learning objectives, strengths, and weaknesses of students). The utilization of personalized educational data mining not only benefits students but also empowers educators and institutions. By leveraging statistical techniques and machine learning methods, educators can gain deeper insights into academic performance, learning preferences, and behavioral patterns of students, enabling tailored learning experiences. This paper aims to demonstrate how data mining methods incorporate personalized education and provide deeper insights closely linked to four primary research fields: educational recommendation, cognitive diagnosis, knowledge tracing, and learning analysis. It stems from analyzing 168 educational data mining papers from related top-tier journals and conferences, among which 149 address these areas (88.7%), which highlights their prominence and frequent exploration in personalized educational data mining research.

Specifically, educational recommendations (ER) are crucial for analyzing learners' preferences and customizing learning materials for individual learners. It involves systems that analyze a learner's past behavior, preferences, and sometimes demographic information to suggest relevant courses, knowledge concepts, or educational resources tailored to meet each learner's unique educational needs. Additionally, cognitive diagnosis (CD) focuses on identifying students' strengths and weaknesses to guide specific instructional interventions. It involves detailed analysis of student's responses to assessment questions and mining knowledge dependencies to evaluate and understand student's cognitive states, including their mastery of various skills or knowledge concepts. Unlike assuming a static knowledge state in CD and providing a snapshot of current cognitive results, knowledge tracing (KT) is a predictive modeling technique in personalized education. It tracks and forecasts learners' knowledge acquisition over time by analyzing their sequential interactions with educational materials. While sharing sequential data with cognitive diagnosis, knowledge tracing stands out due to its predictive focus. It concentrates on forecasting future learning outcomes, as opposed to cognitive diagnosis, which is more oriented toward di-

Zhuang Liu (✉)
E-mail: liuzhuang@buaa.edu.cn

agnosing current cognitive states. In contrast to the aforementioned tasks, learning analysis (LA) in personalized education concentrates on comprehensively analyzing students' behavioral patterns, interaction styles, and learning habits during educational activities.

*Paper collection.* In our study, we first search the related top-tier conferences and journals like The International Conference of World Wide Web (WWW), Web Search and Data Mining (WSDM), Association for Computing Machinery's Special Interest Group on Knowledge Discovery and Data Mining (KDD), The ACM Special Interest Group on Information Retrieval (SIGIR), The Conference on Information and Knowledge Management (CIKM), International Conference on Educational Data Mining (EDM), IEEE Transactions on Knowledge and Data Engineering (TKDE), from 2013 to 2023. We use specific search terms for each field: "educational," "course," "concept," "educational resource" plus "recommendation" for educational recommendation; "cognitive diagnosis" for cognitive diagnosis; "knowledge tracing/tracking" for knowledge tracing; and "learning," "behavior," "predictive" plus "analysis" for learning analysis (Figure 1). Then, we thoroughly examine the citation graph of the identified papers, retaining those that primarily address these areas.

*Related surveys.* Despite abundant research in personalized educational data mining, systematic reviews of these studies are limited. Numerous studies (Dalipi et al., 2018; Kundu et al., 2021; Shristi et al., 2020; Tarus et al., 2018; Thongchotchat et al., 2023; Urduaneta-Ponte et al., 2021) have provided insightful reviews of recommender systems in education. Meanwhile, the most recent surveys (Abdelrahman et al., 2023; Liu et al., 2023) offer systematic overviews of knowledge tracing and cognitive diagnosis, respectively. However, these existing surveys have solely concentrated on specific and limited aspects within the field. Conversely, surveys by Bai et al. (2021) and Lin et al. (2023a) broadly cover multiple scenarios but do not offer a comprehensive overview of all key areas in the field. A detailed comparison between our survey and others is presented in Table 1.

To bridge existing gaps, our work systematically consolidates previous research, providing an integrated and comprehensive overview across educational recommendation, knowledge tracing, cognitive diagnosis, and learning analysis within personalized educational data mining. By illuminating current practices and their impacts, we aim to inspire innovative approaches and pave the way for cultivating a more enriched and effective educational landscape. The key contributions of our survey are summarized in Table 1:

**Table 1**  Comparison of Our Survey and Other Related Surveys

| Survey (Authors) | Year | Domain | | Scope | | | |
|---|---|---|---|---|---|---|---|
| | | Specific | General | ER | KT | CD | LA |
| Tarus et al. | 2018 | ✔ | | ● | ○ | ○ | ○ |
| Dalipi et al. | 2018 | ✔ | | ○ | ○ | ○ | ◖ |
| Shristi et al. | 2020 | ✔ | | ● | ○ | ○ | ○ |
| Kundu et al. | 2021 | ✔ | | ● | ○ | ○ | ○ |
| Urdaneta-Ponte et al. | 2021 | ✔ | | ● | ○ | ○ | ○ |
| Bai et al. | 2021 | | ✔ | ◖ | ○ | ◖ | ◖ |
| Liu et al. | 2023 | ✔ | | ○ | ○ | ● | ○ |
| Thongchotchat et al. | 2023 | ✔ | | ● | ○ | ○ | ○ |
| Abdelrahman et al. | 2023 | ✔ | | ○ | ● | ○ | ○ |
| Lin et al. | 2023 | | ✔ | ● | ● | ◖ | ◖ |
| Ours | | | ✔ | ● | ● | ● | ● |

*Note.* ○ represents "not covered," ◖ represents "partially covered," ● represents "fully covered."

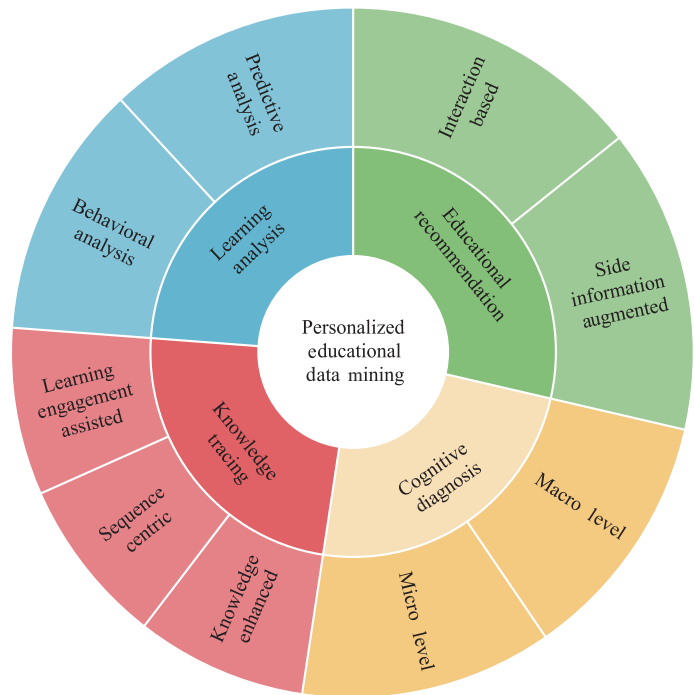

**Figure 1**  Taxonomy of Personalized Educational Data Mining

(1) To the best of our knowledge, our survey stands as the first comprehensive review of recent advanced data mining methods in personalized education. It distinguishes itself by providing a broader and more integrated perspective in general.

(2) We explore four critical areas of personalized educational data mining: educational recommendation, knowledge tracing, cognitive diagnosis, and learning analysis. For each scenario, we present a structural taxonomy and compile the most commonly used datasets, as shown in Figure 1.

(3) Our work identifies and proposes an extensive array of potential future research directions, addressing existing deficiencies in the field and pioneering novel pathways for cutting-edge exploration and innovation.

# 2 Methodology

## 2.1 | Educational Recommendation

In the field of personalized education, it is critical to filter out educational objects to match individual learner profiles adaptively. Aiming to suggest relevant courses, knowledge concepts, or educational resources, educational recommender systems analyze a learner's past behavior, preferences, and sometimes demographic information, as depicted in Figure 2. From a data-centric perspective, personalized education recommender systems are broadly categorized into interaction-based and side information augmented.

### 2.1.1 Interaction-Based Methods

Interaction-based methods aim to recommend candidate lists for students by only focusing on the historical interactions between students and educational objects. Notably, these methods disregard side information and exclusively consider one-hot encodings.

Early research utilized traditional data mining methods to model students' preferences for educational objects. For instance, the massive open online course (MOOC) oriented recommendation system (MCRS) model (Zhang et al., 2018b) employs the Apriori algorithm to recommend suitable courses to students in the Chinese MOOC platform. Goudar and Shidaganti (2023) propose a content-based collaborative filtering method to recommend online courses to students based on their search terms and apply the cosine function to measure the similarity between students and courses. Nguyen et al. (2021) present an MF method based on Funk SVD, resulting in superior performance on the present user- and item-based CF models. To consider the sequential relationship between courses, Polyzou et al. (2019) propose Scholars Walk, a random-walk-based model to recommend a short list of courses for next semester based on students' prior courses.

With the rapid development of deep learning, techniques like neural networks and reinforcement learning (RL) have found extensive applications in uncovering intricate patterns of student interests. Zhang et al. (2019) and Lin et al. (2021) utilize hierarchical RL techniques and dynamic attention to capture user preferences and recommend personalized courses. Lin et al. (2023b) propose a joint learning framework with a multi-scale deep RL method to construct learners' multiple profiles according to coarse-grained and fine-grained semantics. They address the challenges of dilution of multiple preferences caused by attention mechanism but suffer from the trade-off between exploration and exploitation. To address this problem, Lin et al. (2022) incorporate dynamic recurrent mechanisms into hierarchical RL. In addition, Faroughi and Moradi (2022) employ Siamese neural networks (SNNs) to extract latent representations of students and courses to model users' positive and negative preferences.

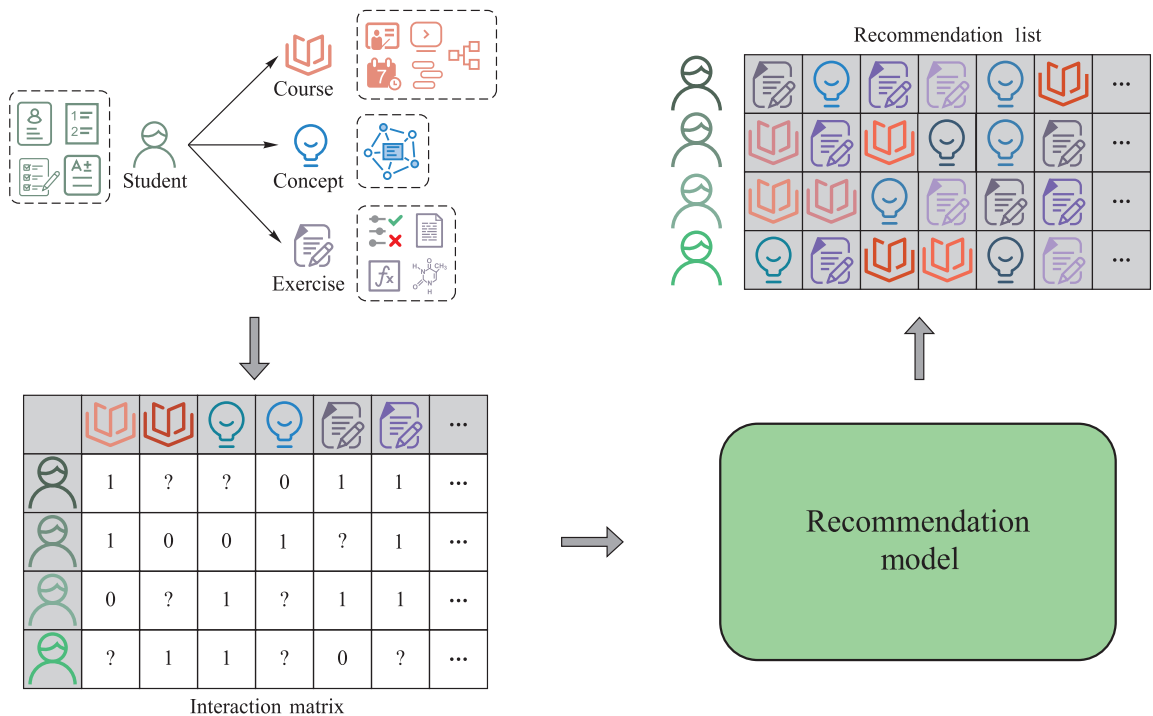

**Figure 2** A Toy Example of an Educational Recommendation

### 2.1.2 Side Information Augmented Methods

Methods based on ID information tend to overlook the diverse side information related to students or educational objects, thus limiting their ability to comprehensively capture students' preferences. To address this limitation, many methods incorporate rich contents into recommender systems and ensure that the recommendations align more closely with the unique educational requirements of each learner. We dive deeper into these methods by the type of recommended objects and relevant side information.

*Courses.* In terms of recommending courses for learners, contents such as course information, concepts associated with the course, and learners' profiles are commonly considered to provide a more accurate depiction of learners' interests. Ma et al. (2017) and Chen et al. (2020) employ TF-IDF and word2vec on course textual information to enhance the similarity measurement between courses. Ma et al. (2020) combine multiple factors, including course popularity, to discern the reasons behind course selections. Additionally, relationships among courses are utilized to identify suitable learning paths (Al-Twijri et al., 2022) and model long-short-term interests (Wang et al., 2022b). Besides, Mondal et al. (2020) conduct K-means clustering by student grades and apply the Apriori algorithm to recommend suitable courses. Instead of considering a single type of content, multi-modal information (Ren et al., 2020) and fused attributes such as sentiment feature & course labels (Ng & Linn, 2017), student grade & course information (Jiang et al., 2019), student profile & course information (El Badrawy & Karypis, 2016), student profile & course relations (Jing & Tang, 2017) and student ratings & course information (Wu et al., 2020a) are also considered.

*Educational resources.* With the success of e-learning platforms, various open educational resources have emerged, defined as teaching, learning, and research materials available in the public domain. Consequently, this abundance has highlighted the challenge of effectively recommending personalized educational resources to learners. Chen et al. (2019) present an adaptation recommendation method based on online learning styles, firstly clustering learners according to their online learning styles and applying item-based CF to recommend relevant learning resources for learners' current needs. To improve the relevance of the recommendations in an online learning context, Baidada et al. (2020) propose a hybrid recommendation method, which combines content-based filtering and collaborative filtering. In e-learning, recommender systems, interaction, and interpersonal information are typically limited. To this end, Ma et al. (2022b) and Wan et al. (2019) propose knowledge graph-based recommendation and learners' influence propagation-based recommendation to alleviate the problem of data sparsity, respectively. Beyond the scope of short-term learning, lifelong learning has also become prominent in the era of e-learning. To build a lifelong learning recommender system, Bulathwela et al. (2020) design a dynamic, scalable, and transparent recommendation method, modeling the learners' knowledge state based on the learning background and novelty of the learning material.

*Knowledge concepts.* The topic of recommending not just courses but specific knowledge concepts and exercises emerges as a crucial area of focus, aiming to fit the learners' personalized learning paths and goals. Liu et al. (2019b) leverage the concept graph of exercises and propose a framework of the combination of recurrent neural network and actor-critic RL algorithm to personalize the learning path. Huo et al. (2020) use a long short term memory (LSTM) network plus a personalization mechanism to represent contextualized information from knowledge concepts. Ai et al. (2019) incorporate exercise labels into dynamic key-value memory network (DKVMN) (Zhang et al., 2017) and build a student simulator with RL to recommend mathematical exercises. Based on the students' exercise answer records, Wu et al. (2020b) propose a method that integrates recurrent neural networks (RNNs) and deep knowledge tracing (DKT) to filter out proper exercises through students' mastery level of knowledge concepts. Combining side information from both students' scores and difficulty of exercises, Huang et al. (2019) design a novel rewards function for multi-objectives with two different Q-networks driven agents. Based on the heterogeneous network constructed by multi-modal information (concepts, teachers, and videos, etc.) and graph neural network with attention mechanism, Gong et al. (2020) and Gong et al. (2023) introduce an extended matrix factorization method and RL to tailor recommendations, respectively.

## 2.2 | Cognitive Diagnosis

In recent years, cognitive diagnosis has garnered significant attention among researchers for its role in identifying students' strengths and weaknesses, enabling targeted instructional interventions. By meticulously analyzing response patterns to assessment questions and precise mining of knowledge dependency, cognitive diagnosis aims to evaluate and understand students' cognitive state at a given time, particularly their mastery of various skills or knowledge areas, illustrated in Figure 3. This paper will present cognitive diagnosis methods from macro and micro levels (Wang et al., 2023a).

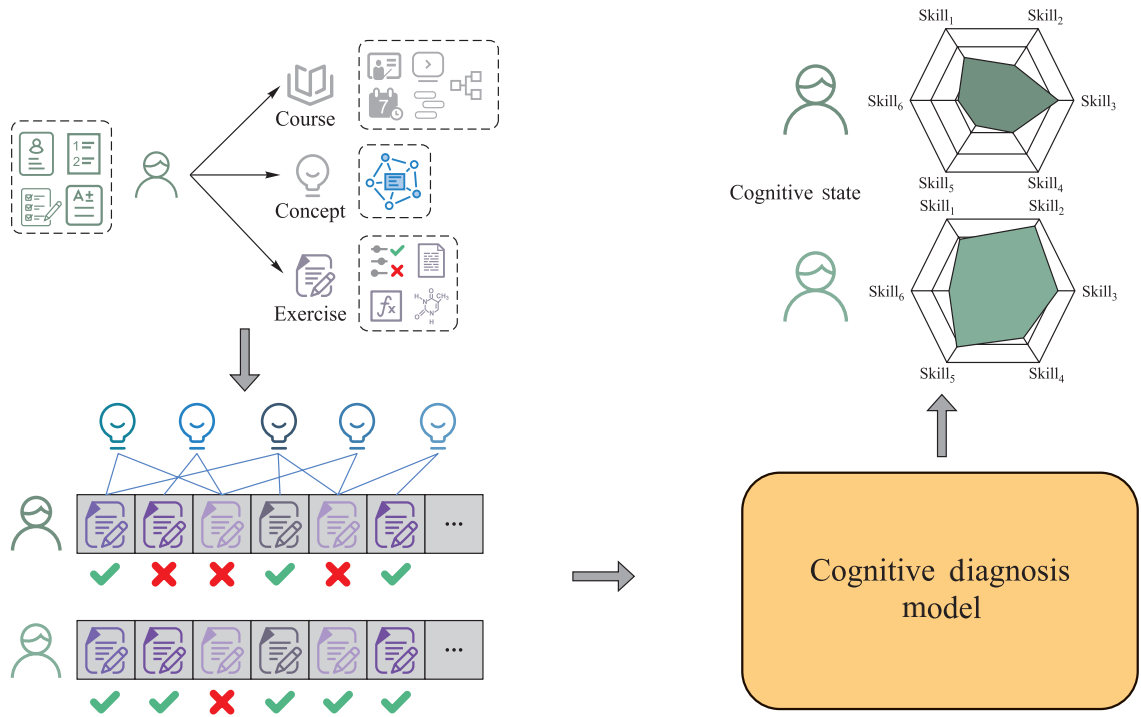

**Figure 3**  A Toy Example of Cognitive Diagnosis

### 2.2.1 Macro Level

The macro-level methods evaluate students' latent skills without considering their specific cognitive attributes. Key examples of this type are classical test theory (CTT) (Crocker & Agina, 1986) and item response theory (IRT) (Rasch, 1993). CTT posits that students' observed test score is a combination of their actual skill level and some error factors. In contrast, IRT determines the probability of a student answering a question correctly through a logistic function, which considers both the students' singular ability and various question-related factors. These factors might include the difficulty level of the question, its capacity to differentiate between varying levels of students abilities and the likelihood of a student guessing the answer correctly.

To improve the interpretability based on psychological theories, Pei et al. (2022) present a self-adaptive attention gate cognitive diagnosis model (AGCDM), which introduces a hierarchical multi-stage architecture that learns the representations of the item response logs, quizzes, and skills and applies self-attention gate mechanisms to capture the rich information and noisy errors in the cognitive diagnosis process. Li et al. (2022) and Bi et al. (2023) introduce Bayesian networks to model the influence of attribute hierarchy and uncertainty on students' cognitive states, respectively. In order to assess the proficiency of a group of students on specific knowledge concepts, Huang et al. (2021) propose a novel framework for multitask-based group-level cognitive diagnosis (MGCD), which adopts a multi-task learning approach to jointly model student-exercise and group-exercise responses and uses a context-aware attention network to aggregate student representations into group representations. Zhou et al. (2021) and Gao et al. (2021), respectively, present a novel framework that induces contexts and structural relations (e.g., students, exercises, and concepts) to improve the CD methods. To address the problem of incremental and non-stationary data in online education systems, Tong et al. (2022b) propose an incremental cognitive diagnosis framework, which consists of a deep trait network (DTN) to acquire the trait parameters of learners and items in an inductive way, and an incremental update algorithm (IUA) to balance the prediction effectiveness and training efficiency with a turning point analysis and a momentum update strategy.

### 2.2.2 Micro Level

Micro level methods focus on conducting detailed diagnoses of students, typically evaluating their competence in each knowledge component. These methods generally fall into two categories: One involves leveraging richer information for analysis, while the other devises innovative diagnosis functions or models.

*Incorporate richer information.* Incorporating various factors, such as exercise factors, guess and slip factors, and knowledge relationships, several studies (Li et al., 2022; Qi et al., 2023; Wang et al., 2020; Wang et al., 2022a; Yang et al., 2022) propose different extendable and general neural network frameworks to provide precise and interpretable diagnosis outcomes, catering to intelligent education systems. With the recent success in graph neural networks, many graph-based CD models are proposed. For instance, Mao et al. (2021) construct a course graph to capture the latent relations of videos and exercises and uses the graph convolutional network (GCN) to learn their representations. Wang et al. (2021c) and Ma et al. (2022a) utilize neural networks to capture information from knowledge concept graphs, to enhance the performance of CD models. Su et al. (2022) present a graph-based cognitive diagnosis model (GCDM), which consists of an attentive knowledge aggregator that selectively gathers information on the heterogeneous graph of students, skills, and questions and a performance-relative propagator that infers the students' cognitive states from their responses. Wu et al. (2022) propose a multi-relational cognitive diagnosis (MRCD) framework, which utilizes the attention mechanism to learn concept-level representations on Q-matrix and uses graph contrastive learning to learn exercise-level representations on views of correct and incorrect patterns, further to diagnose student cognitive states by fusing two level representations. Except for various relational graphs, He et al. (2023) introduce a multihop attention mechanism (MHA) model, using a pre-train language model to embed the mathematical questions, a bi-level LSTM to learn the contextual information, and multi-hop attention to focus on different vital parts of the questions and generate multiple sentence representations.

*Design diagnostic function/model.* Liu et al. (2018) propose a fuzzy cognitive diagnosis framework (FuzzyCDF), which adopts fuzzy set theory to handle the partially correct responses and the skill interactions on different types of problems to model examinees' knowledge state. Zhang et al. (2023a) present a generalized multi-skill aggregation method based on the Sugeno integral (SI-GAM), which also introduces fuzzy measures to model the skill weights and uses various aggregate functions based on the max-min operator of the Sugeno integral to capture the complex interactions and weights of multiple skills. Inspired by the bayesian personalized ranking (BPR) loss function in recommender systems, Tong et al. (2021) propose an item response ranking framework (IRR), which designs an item-specific two-branch sampling method to construct response pairs based on their partial order and uses a pairwise objective function to optimize the monotonicity in the pair formulation. Moreover, Bu et al. (2023) propose a hybrid evolutionary algorithm with a customized local search operator based on a modified probabilistic model, SSVELS, which can detect and correct global collisions in student mastery patterns. Yang et al. (2023) aim to automatically design cognitive diagnosis models by multi-objective evolutionary neural architecture search (NAS). They employed multi-objective genetic programming (MOGP), featuring tailored genetic operations and an initialization strategy to explore the search space to address the challenge of overly simplistic architectures of CDMs, reducing the reliance on human expertise.

## 2.3 | Knowledge Tracing

Adjusting the difficulties and topics of educational content dynamically based on the learners' progress is another essential demand in personalized education. Knowledge tracing is a predictive modeling technique that tracks and forecasts learners' knowledge acquisition over time by analyzing how students interact with course material, as shown in Figure 4. It enables educators to tailor instruction to meet individual learning trajectories. Current knowledge tracing methods mainly focus on modeling learning engagement and knowledge enhancements.

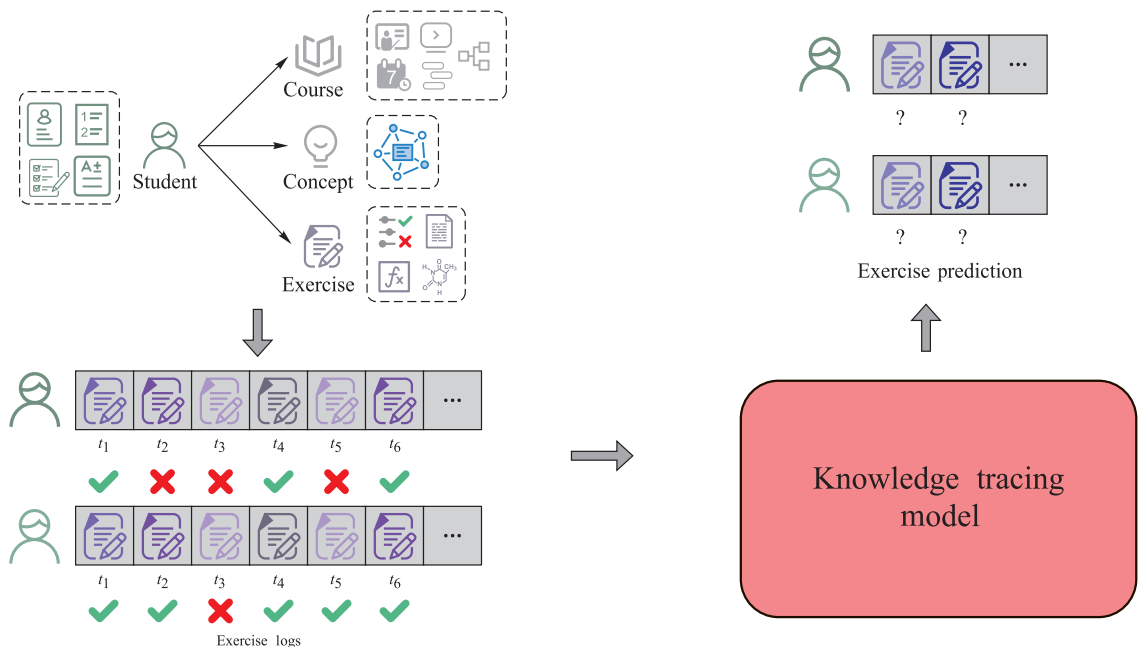

**Figure 4** A Simple Example of Knowledge Tracing

### 2.3.1 Sequence-Centric KT Methods

Sequence-centric KT methods focus solely on modeling the sequence of interactions between students and exercises without considering additional auxiliary information. Traditional KT methods mainly fall into two categories: Bayesian KT (BKT) and factor analysis models. BKT is a hidden Markov model that views each learner's knowledge state as a binary variable and utilizes Bayesian inference to update the state. Typical works include dynamic BKT (Cui et al., 2019) and KT-KDM (Cai et al., 2019). While factor analysis models tend to learn general parameters from historical data for predictive purposes. Vie and Kashima (2019) first apply the item response theory (IRT), additive factor model (AFM), and performance factor analysis (PFA) to learn common factors. Subsequently, factorization machines (FMs) are employed to estimate students' knowledge states.

Recently, numerous researchers have integrated deep neural networks into KT tasks owing to their effectiveness and outstanding performance. The majority of KT methods based on deep learning can be divided into three main types: recurrent neural network (RNN) based, convolutional neural network (CNN) based, and attention (or Transformer) based. These approaches treat a student's exercise records as a sequence and aim to make the next performance prediction along this sequence. For example, Abdelrahman and Wang (2019) design a sequential key-value memory network, which amalgamates the modeling capacities of RNNs with the memory capabilities of existing deep learning KT models to better understand student learning. To consider the need for individualization in student learning, CKT (Shen et al., 2020b) adopts hierarchical convolutional layers to extract individualized learning rates based on continuous learning interactions. While RNN and CNN-based methods have achieved remarkable results, they often lack interpretability and can yield unstable predictions when handling long sequential inputs. To mitigate these challenges, various attention-based KT models (Pandey & Karypis, 2019; Wang et al., 2023b) have emerged. These models assign varying attention weights and directly capture the relationships among each exercise in the input sequence, further alleviating waviness issues. Motivated by the robust sequence modeling capabilities, recent works have turned to the transformer models for KT tasks. Shin et al. (2021) propose a transformer-based knowledge tracing model that integrates two temporal feature embeddings into the response embeddings. However, the sparse interactions between students and exercises pose challenges for the Transformer models, making them prone to overfitting and inaccurate knowledge state capture. In response, Lee et al. (2022) and Yin et al. (2023), respectively, introduce contrastive learning techniques to reveal semantically similar or dissimilar examples of a learning history, promoting a better understanding of their relationships.

### 2.3.2 Learning Engagement Assisted Methods

Learning engagement assisted methods concentrate on the features of students' interactions, encompassing both immediate, short-term factors and enduring, long-term factors. These factors serve as auxiliary information and reflect in-depth characteristics related to students' academic involvement. In doing so, these methods offer a holistic understanding of the diverse elements shaping the educational trajectories of each student.

*Immediate learning engagement.* Immediate factors related to learning enhancement mainly include statistical metrics of students' responses, such as speed and number of attempts. Choi et al. (2020a) present SAINT, leveraging the speed of finishing each question, namely intra-exercise time, as an assistance of sequential exercise data. Shen et al. (2021) introduce intra-exercise and inter-exercise time simultaneously, taking the influence of temporal information into consideration. To further model the learning process with diverse short-term factors, Xu et al. (2023) collect information on speed, attempts, and utilization of hints. Then, a module called fused behavior effect measuring (FEBM) is proposed to capture intricate dependency relations of diverse factors and assist the evolution of students' knowledge status.

*Enduring learning engagement.* Extraction and analysis of long-term information, such as forgetting factors, item response theory (IRT), and inter-sequence signals within exercise sequences, are integral components in modeling students' knowledge evolution dynamics. Incorporating enduring learning engagement is pivotal, as it provides valuable insights into the sustained aspects of students' cognitive development over an extended period. Shen et al. (2021) and Huang et al. (2019) employ sequential information extracted from original exercise sequences. This information includes parameters such as time intervals, knowledge status, and the recurrence of identical questions. The purpose is to systematically calculate the forgetting behaviors exhibited by distinct students within the context of exercise interactions. Meanwhile, other works (Gan et al., 2020; Huang et al., 2020; Zhao et al., 2023b) exhibit a comprehensive perspective by not only accounting for forgetting behavior but also incorporating an emphasis on the learning process. This dual consideration refines the mechanisms inherent in educational scenarios. To further utilize the pedagogical theories, Zhu et al. (2023) and Chen et al. (2023) employ IRT as the foundational scaffold. This strategic utilization of IRT aims to enhance the accuracy of modeling students' cognitive engagement within the exercise sequences. Wang et al. (2021a) delve into the application of the Hawkes Process in probability theory and statistics. This incorporation introduces the theory of causal inference into the field of knowledge tracing, contributing a novel perspective that enhances the understanding of temporal dynamics and their cross-effects within this educational context. Notably, Long et al. (2022) propose CoKT, leveraging inter-student information. This approach involves retrieving sequences from peer students who share similar question-answering experiences. Thus, the inter-student information is seamlessly integrated with intra-student information to effectively trace students' knowledge states and predict the accuracy of their responses to questions.

### 2.3.3 Knowledge Enhanced Methods

Knowledge enhanced methods integrate diverse and intricate side information from exercises such as exercise-concept graphs, question difficulty, and exercise textual content. By leveraging these additional data, knowledge tracing methods deepen the understanding of the exercises, offering a more comprehensive and accurate representation of the learning experience. We classify knowledge-enhanced methods into two types, depending on the types of exercise information used: relation enhanced and textual content enhanced.

*Relation graph enhanced.* Relational graph enhanced methods aim to improve the performance of KT models by leveraging the relational structures between questions and concepts. Tong et al. (2020) propose SKT, exploiting the multiple relations (similarity, prerequisite, etc.) in knowledge structure to propagate the influence among concepts. Zhang et al. (2023b) integrate the hierarchical concept tree with a multi-head attention mechanism into a deep knowledge tracing model. Inspired by the advances in graph neural networks (GNN), Nakagawa et al. (2019) present GKT. Casting the knowledge structure as a graph with various implementations, they reformulate the knowledge tracing task as a time series node-level classification problem in GNN. Yang et al. (2021) propose GIKT, utilizing a graph convolutional network (GCN) to substantially incorporate question-skill correlations via embedding propagation. However, the pairwise structure of GNN neglects the complex high-order and heterogeneous relations among questions and concepts. To address this issue, Jiang et al. (2023) embed a question-concept heterogeneous graph with Metapath2vec (Dong et al., 2017) and design two attention-based encoders to represent the learners' engagements and knowledge states. Tong et al. (2022a) introduce the concept of problem schema and textual information from exercises to construct a hierarchical exercise graph to explore the latent complex relations between exercises. Furthermore, contrastive learning has been incorporated into KT models to improve discriminative ability and robustness. Wu and Ling (2023) develop a novel model with the heterogeneous hypergraph network (HHN) and propose intra- and inter-graph attentions to aggregate information upon HHN, supplied by an auxiliary contrastive learning task. Song et al. (2022) present Bi-CLKT, designing a two-layer comparative learning scheme based on an "exercise-to-exercise" (E2E) relational subgraph, which involves node-level and graph-level contrastive learning with a joint training loss.

*Textual content enhanced.* By extracting rich information from the materials (e.g., explicit labels, exercise texts), textual content enhanced methods achieve superior performance and more interpretable analysis of learner's knowledge acquisition. Shen et al. (2022) propose DIMKT with an adaptive sequential neural network, explicitly incorporating the difficulty into the question representation to establish the relation between students' knowledge state and the question difficulty level. Abdelrahman and Wang (2023) present knowledge augmented data teaching (KADT), developing an attention-pooling mechanism to distill knowledge representations of a student model with respect to class labels under a RL framework. Liu et al. (2020) demonstrate that KT models can realize superior performance by pre-training initialized embeddings for each question on abundant side information (question difficulty and three kinds of relations in the question-skill graph). Liu et al. (2019a) first propose a general exercise-enhanced recurrent neural network by exploring both students' exercise records and contents of corresponding exercises. Liu et al. (2022) conduct the first exploration into open-ended knowledge tracing (OKT) by studying the new task of predicting students' exact open-ended responses to programming questions. They develop a student knowledge-guided code generation approach, combining program synthesis methods using language models with student knowledge tracing methods. To address the limitation of domain-specific and scarcity in most KT methods, Cheng et al. (2022) introduce three-phased AdaptKT, involving instance selection by similar question texts, distribution discrepancy minimization, and output layer fine-tuning to transfer knowledge between domains.

In addition, several methods combine factors from student interactions and extra knowledge contents to improve the performances and achieve a more expressive result. For example, Pandey and Srivastava (2020) propose RKT, introducing a relation-aware self-attention layer with contextual information that integrates both the exercise relation information through their textual content as well as student performance and the forgetting behavior modeled by an exponentially decaying kernel. Abdelrahman and Wang (2022) present a deep graph memory network (DGMN), which captures forgetting behaviors by an attention-based forgetting gated mechanism over the mutual dependencies between concepts and learns relationships between concepts from a dynamic concept graph. Moreover, Xiao et al. (2023) propose knowledge tracing based on multi-feature fusion (KTMFF), which extracts multiple features (the question text, the knowledge point difficulty, the student ability, the duration time, etc.) with the multi-head self-attention mechanism. Li and Wang (2023) propose RAKT, incorporating interaction information (context of exercises and the different time intervals between exercises) and students' behaviors (slipping factor and the guessing factor). Moreover, an extension model QRAKT is developed using a Q-matrix calibration method based on hierarchical knowledge levels to consider the relationship between exercise and knowledge concepts.

## 2.4 | Learning Analysis

In personalized educational systems, it is essential to understand not just what students learn but how they engage with the learning resources. Learning analytics delves into the comprehensive analysis of students' behavioral patterns, interaction styles, and learning habits during educational activities, incorporating not only academic performance but also student participation, time allocation for tasks, and other relevant factors, as shown in Figure 5. This differentiation results in two primary categories of learning analysis methods: behavioral analysis, focusing on patterns during the learning process, and predictive analysis, aimed at forecasting educational outcomes.

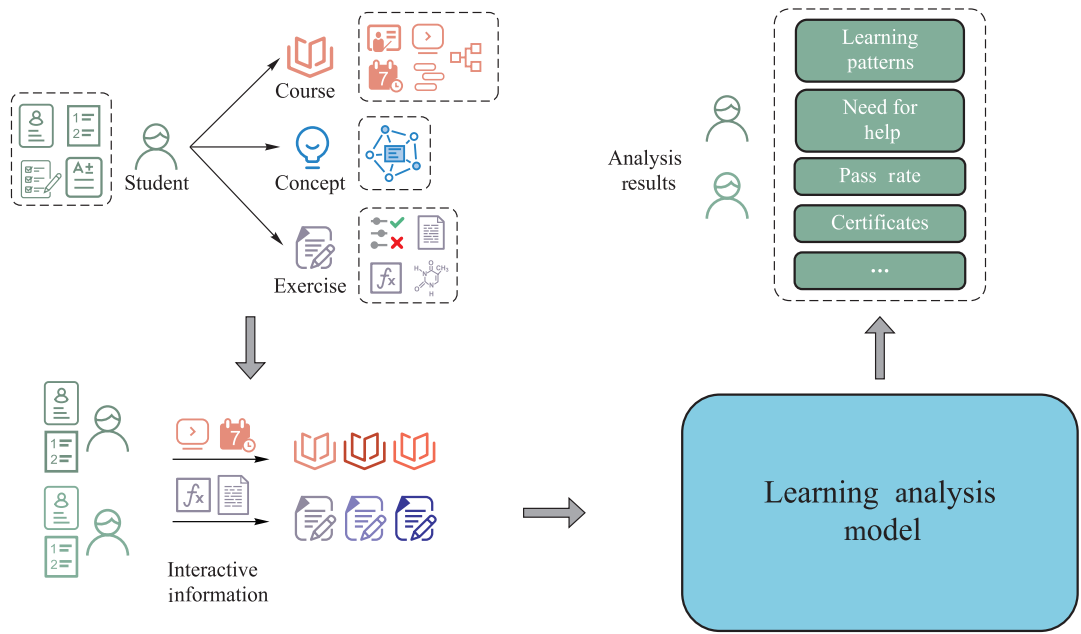

**Figure 5** A Toy Example of Learning Analysis

### 2.4.1 Behavioral Analysis

This section concentrates on the in-depth exploration of student learning patterns and habits during the learning process. It aims to decode the complexities of how students interact with educational materials and engage in course activities. By dissecting these patterns, educators and researchers can gain valuable insights into students' diverse learning styles and strategies. It further aids teachers in adjusting their teaching methods and content delivery, leading to more effective, personalized, and responsive educational experiences.

*Traditional methods.* The first part includes studies that establish the basic principles of behavioral analysis and apply these principles to real-world educational settings, starting with foundational methodologies like sequence mining and linear segmentation. Kinnebrew et al. (2023) present a novel combination of a piecewise linear segmentation algorithm and differential sequence mining to discover different students' cognitive and metacognitive strategies in open-ended tasks. Liu et al. (2017) employ sequential analysis to examine the learning behaviors of students within the cloud classroom platform over a specific period, using sequence analysis to identify key patterns between resources and homework among different grade groups. Han et al. (2017) and Cheng et al. (2018) employ lag sequence analysis (LSA), a pivotal tool, to uncover intricate student behaviors on online educational platforms. Liu et al. (2017) encode the behaviors and extract the two-step lag sequences in learning processes. Then, frequency analysis and sequential analysis are subsequently adopted to discover the distributions and frequency transition patterns of the two-step behavioral sequence.

*Machine learning.* With emerging machine learning techniques, researchers started incorporating advanced methods into behavioral analysis. Zhang et al. (2018a) analyze data from 711 students using K-means and hierarchical clustering algorithms, identifying different learner types (weak cognition, self-consciousness, short-cut, lazy) based on online behaviors and various engagement indicators in order to suggest adjustments. Boroujeni and Dillenbourg (2018) present a hypothe-sis-driven approach to extract predefined patterns (e.g., whether the learners start their learning sequence by watching a video or not) and a data-driven clustering approach for discovering patterns between performances and help requests in an unsupervised manner from MOOC activity sequences. Yan and Au (2019) utilize neural networks to mine online learning behaviors from data such as age, gender, and online engagement, highlighting the significance of days of access and hits count as key predictors of academic performance. Wang et al. (2019) employ a novel approach by using detailed access trajectories (DATs), which are two-dimensional matrices that represent students' interactions with MOOC content over time, and conducted several empirical studies to analyze these detailed trajectories and extract meaningful patterns of student behavior. To investigate the relationship between learning design and learner behaviors, Shen et al. (2020a) employ Bloom's Taxonomy to categorize learning resources and analyze learners' behaviors through visualizations and social network analysis, revealing patterns in resource access and usage. Zhang et al. (2020a) employ a deep belief network (DBN) to classify learning styles into 16 categories with a model based on expert experience and various learning indicators. Zhao et al. (2023a) propose a novel framework that formulates the student simulation task as a Markov Decision Process and uses two-stage imitation learning to model the intentions and behaviors of students, addressing the challenges of exposure bias, single-step optimization, and implicit intentions in existing simulators.

### 2.4.2 Predictive Analysis

In a complementary manner to behavioral analysis, predictive analysis focuses on forecasting future student behaviors and outcomes based on their past and current engagement in educational systems. Unlike knowledge tracing, which primarily tracks and predicts students' knowledge acquisition over time, predictive analysis extends its gaze beyond the immediate learning process to encompass post-course behaviors, such as the likelihood of a student withdrawing from a course, their anticipated academic achievements, or the probability of attaining certification. We will briefly introduce these methodologies in two main directions: One focuses on predicting student dropout rates, and the other on forecasting student performance outcomes.

*Dropout prediction.* The field of dropout prediction in MOOCs focuses on mitigating high dropout rates and enabling educators to provide timely interventions. Halawa et al. (2014) extract various features of learners' study activities, self-proposed, crowd-proposed, and study habits related, and then build logistic regression models to predict the dropout risk and to reduce the high dropout rates by delivering timely intervention. Xing and Du (2019) design a deep learning based weekly temporal dropout prediction model and generate individual student dropout probabilities, which outperforms KNN, SVM, and decision trees in accuracy and personalization. Additionally, Wang and Wang (2019) extract 31 features from learners' activity data and study habits data and use logistic regression to build predictive models for different groups of learners based on their participation level. Feng et al. (2019) conduct a systematic analysis of users' learning activities and context information on two large datasets from XuetangX and propose a context-aware feature interaction network (CFIN) that incorporates context-smoothing and attention mechanisms to predict dropout probability. Goel and Goyal (2020) introduce a semi-supervised learning model that uses both click-stream features and the influence of friends to predict potential dropouts.

*Performance prediction.* In addition to dropout rate studies, significant research focuses on student performance. Qiu et al. (2016) present a novel model called the latent action dynamic factor graph (LadFG), which captures the latent interactions between students, their forum activities, and their learning behaviors, over time and delves into the factors that influence students' engagement and performance in MOOCs. Wang et al. (2021b) introduce a novel deep learning framework to improve the student learning outcome prediction task, which simulates the cognitive process of students in answering questions and integrates both question explanations and student responses as constraints for supplementary supervision. Some studies specifically concentrate on analyzing students' performance in coding tasks, offering valuable insights into their problem-solving abilities and learning progression. Yan et al. (2017) examine interaction logs from a programming game and extract a set of features related to learner engagement, commitment, and behavior, then develop three traditional machine learning classifiers for each level of the game. Mao (2019) proposes a recent temporal pattern (RTP) mining approach that can extract interpretable and meaningful temporal patterns from student interaction data to predict student success and difficulty during a novice programming task. Malysheva and Kelleher (2020) use a bug taxonomy to classify the types of bugs in student code and train random forest models to predict the number of tests to fix the bugs and the chance of abandonment using features derived from the bug types and temporal patterns, to help teachers optimize their time and attention. El Aouifi et al. (2021) collect and classify learners' clicks data from four video courses on C++ language and apply K-nearest Neighbors and MLP to predict learners' performance based on their video sequences viewing behavior.

There is also a growing body of work combining both behavioral and predictive analysis to offer a more comprehensive understanding of student learning processes and outcomes. Worsley et al. (2015) explore the use of affect- and pose-based segmentation as alterna-tives to human-based and fixed-window segmentation for analyzing data from a hands-on engineering design task, comparing the ability to correlate and predict three objectives: success, learning, and experimental condition. Shi et al. (2017) present a non-parametric Bayesian model that captures the homogeneity and heterogeneity of learning behaviors by clustering them into latent student groups, each characterized by a Markov model and different distributions. The model is evaluated by predicting student retention, course completion, satisfaction, and demographics. Zhang et al. (2017b) introduce 19 behavior indicators that cover the whole online learning process and use correlation analysis and logistic regression to select the relevant indicators and build the prediction model. Sedrakyan et al. (2020) propose a conceptual model for designing learning analytics dashboards that provide process-oriented feedback, support diverse learning goals, and enhance cognitive and behavioral aspects of learning. Lu et al. (2018) apply learning analytics for various learning activities (e.g., video-viewing, out-of-class practice, homework, quiz, and after-school tutoring) in a Calculus course and uses principal component regression to predict students' final academic performance, further to provide timely interventions for at-risk students.

# 3 Datasets

To provide a more thorough overview of frequently used public datasets in educational data mining, we have compiled them, showcased in Appendix A. Notable for their extensive use in numerous studies, these datasets play an essential role in progressing the research. We provide key details for each dataset in the table, such as its name, source population, description of data format, subject, URL, and applied scenarios.

### 3.1 │ EdNet

EdNet (Choi et al., 2020b), a dataset from Santa, an AI tutoring platform in Republic of Korea, encapsulates two years of interactions from approximately 780,000 students. It uniquely includes data across multiple platforms (iOS, Android, Web) and is structured hierarchically into four detailed subsets (KT1–KT4). This dataset stands out for its rich representation of student learning behaviors, spanning video lectures, problem-solving, and expert commentaries.

### 3.2 │ XuetangX

The XuetangX dataset (Feng et al., 2019), derived from one of China's largest MOOC platforms, includes ex-tensive data from over 1,000 courses across twelve categories, such as art, computer science, and engineering. The dataset records user activities over specific periods, detailing each activity event and its timestamp within user sessions in various courses, user profile (e.g., user ID, gender, education level, and birth year), and course details (e.g., course ID, start-end dates, course type, and category).

### 3.3 │ MOOCCube

MOOCCube (Yu et al., 2020), a large-scale data repository, encompasses over 700 MOOC courses, 100,000 concepts, and 8 million student behaviors enriched with external resources. It integrates courses, concepts, student interactions, and relationships, facilitating diverse educational data mining scenarios.

### 3.4 │ MOOCCubeX

MOOCCubeX (Yu et al., 2021) is an extensive dataset of MOOCCube with 4,216 courses, 230,263 videos, 358,265 exercises, and 637,572 concepts, alongside over 296 million student behavioral records. It offers a comprehensive, fine-grained concept graph for adaptive learning, integrating vast internal and external MOOC resources to support diverse educational research.

### 3.5 │ MoocRadar

Integrated from MOOCCube, MoocRadar (Yu et al., 2023) is a rich educational dataset featuring 2,513 exercises, 5,600 knowledge concepts, and over 12 million behavioral records. It emphasizes exercise-centric data organization and expert annotation, serving as a valuable resource for knowledge tracing and cognitive diagnosis research.

### 3.6 │ ASSISTments

The ASSISTments datasets from an online tutoring system comprise skill builder problem sets for knowledge practice. They include detailed student interactions, accurate labels, and concept clarity and vary in student count and data completeness.

### 3.7 │ Junyi

The Junyi Academy Online Learning Activity Dataset[1] encompasses over 16 million exercise attempt logs from more than 72,000 K-12 students. It systematically includes detailed records of learning activities and responses, along with a rich assortment of learning behavior characteristics and auxiliary information.

---

[1]  https://www.kaggle.com/datasets/junyiacademy

### 3.8 | Eedi

The Eedi dataset comprises K-12 student responses to mathematics questions (Wang et al., 2021d) collected from Eedi, a globally renowned educational platform used by millions of students. Eedi provides diagnostic questions primarily in mathematics, and each question in the dataset is formatted as a multiple-choice query with four options, where only one choice is correct.

### 3.9 | Math1&2

The two datasets[2] are compiled from final mathematics examinations taken by high school students. They encompass 40 questions, represented by 23 distinct mathematical concepts, and individual performance on each question from 8,813 students.

### 3.10 | OLI Engineering Statics

The dataset originates from a college-level engineering course titled "OLI Engineering Statics 2011" at Carnegie Mellon University[3]. It encompasses a comprehensive collection of 189,297 interactions involving 333 students who engaged with 1,223 distinct concepts within the course.

### 3.11 | Slepemapy.cz

This dataset originates from an online adaptive system (Papoušek et al., 2016), slepemapy.cz, providing adaptive practice of geography facts like names of cities and mountains. It comprises statistics including 91,331 students, 1,459 geographical items, and 10,087,306 answers in total.

### 3.12 | KDD Cup

The KDD Cup dataset comes from the KDD Cup 2010 Educational Data Mining Challenge[4]. It is structured by logs of the interactions between students and computer-aided tutoring systems, focusing on the process of solving algebra problems. The key terms of the dataset include student, problems, step, metrics indicating performances on the step (e.g., incorrections, hints, error rate), and knowledge components.

## 4 Future Work

In this section, we intensively concentrate on the advancing frontiers of personalized educational data mining. Firstly, we will explore the future directions of the broader field of personalized educational data mining, focusing on emerging trends and advancements. Subsequently, we will narrow our focus to specific scenarios, presenting unique challenges and opportunities.

### 4.1 | Explainability

In the advanced landscape of personalized educational data mining, the integration of explainability and interpretability is set to enlighten future research. Its potential lies in the ability to provide transparency and understanding in the decision-making processes of personalized education systems, which is crucial for the stakeholders (students, educators, and policymakers), as education prioritizes understanding the scientific principles and causal relationships of things. For instance, an explainable recommendation system can justify its suggestions for algebra problems by indicating that a student has already mastered the knowledge in other chapters or shows a higher interest in algebra problems (Bao et al., 2023). Cognitive diagnosis (Yang et al., 2022) and knowledge tracing (Shen et al., 2022) models can provide explanations for assessing students' knowledge states by reasoning on the paths in the exercise-concept knowledge graph. In learning analysis, explainable models transcend the limitations of black-box modeling by providing insights into the causal relationships between students' behavioral patterns (Mao, 2019) and their learning outcomes, thereby enhancing the depth and interpretability of the analysis.

### 4.2 | Multimodal Learning

The incorporation of multimodal learning in personalized educational data mining is limited (Ren et al., 2022) and offers a broad landscape for future research. The challenge lies in effectively utilizing and modeling the rich, multimodal information present in educational resources, such as images and audio, to enhance the educational process. For instance, a system could analyze students' interaction with images (picture storytelling) in exercises and recommend similar or more challenging practices in elementary Chinese language courses. In English courses, listening tests require students to answer questions based on the audio they hear. How to model the characteristics of these exercises using multimodal representation learning techniques, enabling cognitive diagnosis models to diagnose students' cognitive states and knowledge tracing models to more accurately track students' performance on such types of questions, is a challenging and under-explored issue. Learning analysis could involve analyzing patterns in how different types of learners interact with and respond to various media, thereby offering deeper insights

---

into learning behaviors and outcomes.

### 4.3 │ LLMs for Personalized Educational Data Mining

Recently, the field of personalized educational data mining has been revitalized by the emergence of generative models, particularly large language models (LLMs). Their profound understanding and generative capabilities make them invaluable tools for this domain. For instance, many recommendation systems based on LLMs use these models not only to generate enhanced data but also to provide rational explanations for their recommendations (Bao et al., 2023). These systems can be adapted to various datasets and educational contexts with simple fine-tuning. Moreover, the ability of large language models to understand long text enables automatic scoring and feedback for student essays, oral presentations, and programming codes. This expands the application scenarios for knowledge tracing and cognitive diagnosis models, significantly reducing teachers' workload. Additionally, large language models are instrumental in developing more advanced and comprehensive Intelligent Tutoring Systems (ITS) (Mousavinasab et al., 2021) and chatbots (Dan et al., 2023). These systems can understand and respond to student needs, aiding teachers in addressing the reluctance of some students to communicate face-to-face.

### 4.4 │ Uncertainty Modeling and Quantification

In educational contexts, numerous factors influence a student's state and performance, many of which are challenging to collect or quantify through data. Additionally, due to privacy concerns, complete access to a student's information is often unattainable. This leads to inherent uncertainties in model predictions, and overlooking these uncertainties can misguide students, potentially resulting in irresponsible teaching practices (Bi et al., 2023). Therefore, incorporating uncertainty into personalized educational data mining methods is essential. One of the classic methods to model uncertainty in deep learning is Bayesian neural networks (Abdar et al., 2021; Kendall & Gal, 2017). Some work in cognitive diagnosis models has already started to integrate Bayesian modeling approaches (Bi et al., 2023). However, introducing methods capable of modeling and quantifying uncertainty into other scenarios, such as recommendation and learning analysis, remains a promising yet challenging task.

### 4.5 │ Datasets and Metrics

After a thorough understanding of commonly used datasets in Section 3, it becomes evident that future de-velopment hinges on creating inclusive datasets. In light of this, we suggest two possible directions in dataset development: the construction of multimodal datasets and the establishment of datasets for diverse subjects. Only a limited number of studies have focused on multimodal educational data mining, primarily due to the scarcity of publicly available datasets containing multimodal data. Student exercises frequently incorporate multimodal elements like images (e.g., topographic maps in geography, molecular structures in biology) and audio (e.g., English listening comprehension). The construction of such multimodal datasets can robustly support models employing multimodal representation learning techniques. On the other hand, the datasets predominantly originate from mathematics, with a notable deficiency in different subjects. There is a need to develop additional datasets for various subjects (such as physics, chemistry, history). These datasets should encompass an adequate number of exercises, concept details, and student response records, similar to those from mathematics, to enable the effective application of existing educational data mining methodologies.

Besides, personalized education for students often involves a variety of goals that are difficult to quantify. Customizing learning plans for students is influenced by multiple factors, such as their knowledge state, the difficulty of the tasks, and their learning habits. However, current research primarily focuses on whether students can correctly solve problems or pass courses as the primary metrics for model evaluation. This singular metric leads to a limited and partial understanding of student states and learning goals. The challenge lies in quantifying these diverse learning objectives or establishing a unified standard to accurately reflect students' educational needs.

### 4.6 │ Open-Ended Questions in Knowledge Tracing

An open-ended question is a type of question that cannot be answered with a binary-value or with a limited set of options. Instead, it requires a more detailed, descriptive, or explanatory response. Writing and comprehensive reading in K-12 education are two typical open-ended questions. While most KT models emphasize response correctness analysis, they often overlook the information embedded in the specific content of responses to open-ended questions. Existing studies (Liu et al., 2022) primarily concentrate on programming questions for computer science education, leaving a substantial gap in exploring open-ended questions in K-12 education.

### 4.7 │ Domain Adaptation

The current state of personalized educational data mining methods, largely domain-specific, presents a limita-tion in the diverse and multifaceted world of education, which encompasses domains ranging from mathematics to geography. The challenge posed by insufficient data in specific domains, as highlighted in the referenced dataset statistics, necessitates innovative solutions. For example, a system trained in mathematics could tailor its recommendations for students, and problem-solving cognitive skills applied in Physics might find parallels in engineering, albeit within distinct content domains. Existing efforts like AdaptKT (Cheng et al., 2022) indicate the potential of using knowledge from data-rich domains to augment models in data-scarce domains, which requires understanding and mapping the structural similarities and differences between these domains.

### 4.8 | Incremental Learning

With the continuous update of data and the evolving learning states of students, Incremental Learning is crucial but rarely explored by limited work (Tong et al., 2022). An incremental learning approach enables educational models to continuously adapt and update in response to real-time changes, ensuring that content in online platforms remains relevant and aligned with the students' evolving knowledge levels. Additionally, Incremental Learning significantly reduces the costs associated with model training and alleviates the workload on online education platforms. Updating models with new data without requiring complete retraining presents a more efficient and resource-effective approach to maintaining and improving educational systems. Besides, Incremental Learning is also vital for bridging the knowledge gap between different grades. It ensures a seamless educational progression, taking into account the cumulative knowledge students acquire over educational periods and adjusting content to their evolving understanding and educational stages.

## 5  Conclusion

The fusion of digital technology and education is becoming more comprehensive and in-depth, making digitalization a key field in modern educational reform and innovation. This involves the expansion of digital technology applications in education, enhancing the digital education system, and strengthening the construction of digital resources. Innovations in education leveraging new digital technology features continue to deepen. Rapid advancements in AI have led to the evolution of AI integration in education across three paradigms: AI-directed (learner-as-recipient), AI-supported (learner-as-collaborator), and AI-empowered (learner-as-leader) (Ouyang & Jiao, 2021). Personalized educational data mining methods add new momentum to the latter two paradigms, which can precisely profile individual learning behaviors, recommending learning resources and paths tailored to the learners' states, thus providing personalized education services. It also enables comprehensive analysis and evaluation of learners' psychological states, effectively addressing teaching feedback issues, and transforming educational evaluation.

In this work, we comprehensively survey recent advancements in data mining methods for personalized education, covering four critical areas: educational recommendation, cognitive diagnosis, knowledge tracing, and learning analysis. It uniquely offers a broad, integrated view, addressing gaps in previous surveys. The paper introduces structured taxonomies for each area, compiles common datasets, and outlines several future research directions, thereby pioneering innovative pathways for exploration and innovation in personalized educational data mining.

**Acknowledgments** This work is supported by the National Natural Science Foundation of China (No. 62377002). We would like to thank Weijie He, Chenyang Lei, Keqin Peng, Keyi Dai, Zibin Zhao, Tong Chi, Shikang Bao, Guanming Chen for their contributions to this work, as they have provided significant assistance in the collection and compilation of existing literature.

# Appendixes

## Appendix A
### Compilation of Commonly Used Datasets

| Datasets | Source | Description of data format | Subjects | URL | Scenarios |
|---|---|---|---|---|---|
| EdNet | Both | EdNet comprises detailed actions like video lecture views, problem-solving, and expert commentary engagement. It includes data on learning material, time allocation, and user behavior. | English | URL | ER, KT, BA, CD |
| XuetangX | Higher education | The XuetangX dataset includes a comprehensive collection of user activities, user profiles, and course information from XuetangX. | Diversity | URL | ER, BA |
| MOOCCube | Higher education | MOOCCube includes over 700 courses, detailed video subtitles, teacher and organization descriptions, concepts with prerequisite chains, and over 190,000 student behavior data. | Diversity | URL | ER, CD, BA |
| MOOCCubeX | Higher education | MOOCCubeX encompasses 4,216 courses, 230,263 videos, 358,265 exercises, 637,572 concepts, and 296 million student behavior data. It also integrates fine-grained concept graphs. | Diversity | URL | ER, CD, BA |
| MoocRadar | Higher education | The dataset includes 2,513 exercises, 5,600 knowledge concepts, and 12 million behavioral records. | Diversity | URL | ER, KT, CD, BA |
| ASSISTments | K-12 | ASSISTments encompasses detailed student interaction records, problem-solving attempts, and knowledge component (KC) labels, with varying complexity, student performance data, and concept-specific learning analytics. | Math | URL | ER, KT, CD, BA |
| Junyi | K-12 | Junyi Academy contains over 16 million problem attempts by 72,000+ students, detailed exercise metadata, and student demographics. | Diversity | URL | ER, KT, CD |
| KDD Cup | K-12 | KDD Cup consists of detailed records of interactions between students and tutoring systems, containing key terms, like problems, steps, knowledge components and opportunities. | Math | URL | ER, KT, CD, BA |
| Eedi | K-12 | Eedi comprises K-12 student responses to mathematics questions, formatted as multiple-choice queries, containing records from 118,971 students, 27,613 questions and 15,867,850 answers. | Math | URL | KT, CD |
| Math1&2 | K-12 | These two datasets are collected from two final mathematical exams from high school students, including problems, concepts, and performances from students. | Math | URL | KT, CD |
| Engineering Statics | Higher education | It is a course-specific dataset from CMU, containing 189,297 interactions between 333 students on 1,223 concepts. | Engineering | URL | ER, KT, CD |
| slepemapy.cz | Both | The dataset contains logs of practices of geography facts from an online system, such as user identifiers, asked and answered place identifiers, question type, and response time. | Geography | URL | KT |

## Appendix B
### Summary of Educational Recommendation Methods

| Method | Year | Venue | Object | Techniques | Side information |
|---|---|---|---|---|---|
| MCRS | 2017 | ICAT2E | Course | Apriori | – |
| Scholars walk | 2019 | EDM | Course | MDP | – |
| HRL | 2020 | AAAI | Course | Hierarchical RL | – |
| Nguyen et al. | 2021 | *Education and Information Technologies* | Course | Cosine similarity, Pearson correlation, funk SVD | – |
| DARL | 2021 | *Knowledge-Based Systems* | Course | MDPs(Markov Decision Processes), RL, attention | – |
| HELAR_W | 2022 | *Knowledge-Based Systems* | Course | RL | – |
| SMRC | 2022 | ICeLeT | Course | Siamese neural network | – |
| Goudar & Shidaganti | 2023 | NMITCON | Course | MF | – |
| RPPR | 2023 | *Applied Soft Computing* | Course | RL, attention MLP | – |
| MEUR | 2022 | *Education and Information Technologies* | Course | RL, GCN | Concept graph |

| Method | Year | Venue | Object | Techniques | Side information |
|---|---|---|---|---|---|
| FKGCF | 2022 | *Computational Intelligence and Neuroscience* | Course | TransE, CF | Concept graph |
| Ma et al. | 2017 | CCSSE | Course | Doc2vec, LSI, TF-IDF | Course information |
| HCR | 2020 | EDM | Course | MF | Course information |
| ItemCF-TFIDF | 2020 | ICBDIE | Course | MF, TF-IDF | Course information |
| Chang et al. | 2016 | Algorithms | Course | AIS cluster | Course ratings, student grades |
| (ES)^2P | 2022 | *Cognitive Computation* | Course | Evolutionary search | Course relation |
| HGNN | 2022 | *Information Processing & Management* | Course | GNN, attention, RNN | Course relation |
| Ren et al. | 2022 | *Sustainability* | Course | LSTM, attention | Multi-modal |
| CrsRecs | 2017 | IISA | Course | LDA, SentiWordNet, MF, MLP | Sentiment feature, course labels |
| Mondal et al. | 2020 | ICCSEA | Course | K-means, MF, Apriori | Student grade |
| Jiang et al. | 2019 | LAK | Course | LSTM | Student grade, course information |
| El Badrawy &Karypis | 2016 | RecSys | Course | MF | Student profile, course information |
| HCACR | 2018 | WI | Course | MF, LDA | Student profile, course relations |
| Wu et al. | 2020a | *Journal of Physics: Conference Series* | Course | MF | Student rating, project attribute |
| Fang et al. | 2021 | Complexity | Edu resource | MF | – |
| Ma et al. | 2022b | ICETC | Edu resource | TransD, matrix factorization | Concept graph |
| TrueLearn | 2020 | AAAI | Edu resource | Bayesian model, concept extraction, IRT | Knowledge content |
| AROLS | 2019 | *Tsinghua Science and Technology* | Edu resource | K-means, Apriori, MF | Learning style |
| SI-IFL | 2020 | TKDE | Edu resource | Active learning, self-organization theory, fuzzy logic | Student profile |
| Baidada et al. | 2020 | *IADIS International Journal on WWW* | Edu resource | Content-based CF | Student profile |
| CSEAL | 2019 | KDD | Knowledge concept | Adaptive learning, LSTM, RL | Concept graph |
| LSTMCQ, LSTMCQP | 2020 | *Information Science* | Knowledge concept | LSTM | Exercise content |
| DKVMN-CA | 2019 | EDM | Knowledge concept | RL, DKVMN | Exercise label |
| ACKRec | 2020 | SIGIR | Knowledge concept | GCN, attention, HIN | Multi-modal |
| HinCRec | 2023 | *Transactions on the Web* | Knowledge concept | HIN, RL | Multi-modal |
| KCP-ER | 2020 | KBS | Knowledge concept | LSTM, DKT | Student response |
| DRE | 2019 | CIKM | Knowledge concept | RL, GRU, Bi-LSTM | Student response, exercise information |

## Appendix C
### Summary of Cognitive Diagnosis Methods

| Method | Year | Venue | Techniques | Category |
|---|---|---|---|---|
| MGCD | 2021 | ICDM | Attention | Macro |
| ECD | 2021 | KDD | Hierarchical attention | Macro |
| RCD | 2021 | SIGIR | Hierarchical attention | Macro |
| ICD | 2022 | KDD | Inductive learning, incremental learning | Macro |

| Method | Year | Venue | Techniques | Category |
|---|---|---|---|---|
| AGCDM | 2022 | ICDM | Attention, GRU | Macro |
| BETA-CD | 2023 | AAAI | Bayesian network, meta-learning | Macro |
| FuzzyCDF | 2018 | TIST | DINA, MCMC | Micro |
| NeuralCDM | 2020 | AAAI | MLP, CNN | Micro |
| IRR | 2021 | IJCAI | Pairwise sampling | Micro |
| LCD | 2021 | ICPCSEE | BERT, GCN | Micro |
| CDGK | 2021 | CIKM | MLP, IRT | Micro |
| PAKP | 2022 | CIKM | MLP, attention | Micro |
| FuzzyCDF-SI-GAM | 2022 | WWW | Fuzzy measure, Sugeno integral | Micro |
| QRCDM | 2022 | *Expert Systems with Applications* | Alpha-partition, MF | Micro |
| GCDM | 2022 | *Knowledge-Based Systems* | Hierarchical graph attention network | Micro |
| HierCDF | 2022 | KDD | Bayesian network, monotonic perceptron | Micro |
| ICD | 2022 | *Expert Systems with Applications* | alpha-partition | Micro |
| KSCD | 2022 | CIKM | MLP | Micro |
| MRCD | 2022 | CICAI | GCN, self-attention | Micro |
| NeuralNCD | 2022 | *Applied Sciences* | MLP, IRT | Micro |
| NeuralCD | 2022 | TKDE | CNN, MLP | Micro |
| EMO-NAS-CD | 2023 | arxiv | Neural architecture search (NAS), evolutionary algorithm | Micro |
| He et al. | 2023 | *Applied Intelligence* | Multihop attention, Bi-LSTM, ALBERT | Micro |
| MHGA_CDM | 2023 | GECCO | Evolutionary optimization | Micro |

## Appendix D
### Summary of Knowledge Tracing Methods

| Method | Year | Venue | Techniques | Category | Interaction Patterns | Knowledge Enhanced |
|---|---|---|---|---|---|---|
| SAKT | 2019 | EDM | Attention | Sequence centric | – / Attention weighted | – |
| Cui et al. | 2019 | JEDM | BKT, Dynamic BKT | Sequence centric | – / Bayesian | – |
| KTM | 2019 | AAAI | IRT | Sequence centric | – / Factorization machine | – |
| SKVMN | 2019 | SIGIR | Attention, LSTM | Sequence centric | – / KV memory | – |
| KT-KDM | 2019 | ICCC | BKT, A2C | Sequence centric | – /RL | – |
| AKT | 2020 | SIGKDD | Attention, Seq2Seq | Sequence centric | – / Attention weighted | – |
| CKT | 2020 | SIGIR | CNN | Sequence centric | – / CNN | – |
| SAINT+ | 2021 | LAK | Attention, transformer | Sequence centric | – / Transformer | – |
| CL4KT | 2022 | WWW | CL, transformer | Sequence centric | – / Contrastive learning | – |
| GameDKT | 2022 | *Expert Systems with Applications* | CNN, RNN, LSTM | Sequence centric | – / New scenario | – |
| KTA, KTA_LSTM | 2022 | *Applied Intelligence* | Attention, LSTM | Sequence centric | – / Finite state automaton | – |
| DTransformer | 2023 | WWW | CL, transformer | Sequence centric | – | – |
| Nagatani et al. | 2019 | WWW | RNN | Learning engagement assisted | Forgetting behavior | – |
| KPT | 2020 | TOIS | GRU | Learning engagement assisted | Learning curve + forgetting curve | – |
| KTM-DLF | 2020 | *Applied Intelligence* | NN | Learning engagement assisted | Item difficulty + learning curve + forgetting curve | – |
| SAINT | 2020 | L@S | Attention, transformer | Learning engagement assisted | Speed / transformer | – |
| LRKT | 2021 | KDD | GRU, MLP | Learning engagement assisted | – / Forgetting gated RNN | – |

*(Continued)*

| Method | Year | Venue | Techniques | Category | Interaction Patterns | Knowledge Enhanced |
|--------|------|-------|-----------|----------|---------------------|-------------------|
| HawkesKT | 2021 | WSDM | Hawkes Process, resampling strategy | Learning engagement assisted | Hawkes Processing | – |
| CoKT | 2022 | WSDM | RNN | Learning engagement assisted | Collaborative signals in students | – |
| TCKT-FI | 2023 | ICCECT | Temporal convolutional network | Learning engagement assisted | Forgetting factors + IRT | – |
| GFLDKT | 2023 | *Information Processing & Management* | GRU | Learning engagement assisted | Learning behavior + forgetting behavior | – |
| QIKT | 2023 | AAAI | LSTM, item response theory（IRT） | Learning engagement assisted | IRT | – |
| LBKT | 2023 | KDD | Tensor fusion network | Learning engagement assisted | Speed + attempts + hints + forgetting factor | – |
| EKT | 2019 | TKDE | LSTM, attention | Knowledge enhanced | – | Textual content |
| GKT | 2019 | WIC | GNN | Knowledge enhanced | – | Knowledge concept graph |
| GIKT | 2020 | ECML PKDD | GCN, RNN | Knowledge enhanced | – | Question-student-skill graph |
| SKT | 2020 | ICDM | GRU | Knowledge enhanced | – | Knowledge relations |
| PEBG | 2021 | IJCAI | Bipartite graph, MLP | Knowledge enhanced | – | Question difficulty + question–skill graph |
| KADT | 2021 | KBS | SKVMN, DKT, LSTM, RL | Knowledge enhanced | – | Course label |
| AdaptKT | 2022 | WSDM | LSTM, transer learning | Knowledge enhanced | – | Textual content |
| DIMKT | 2022 | SIGIR | Sequential NN | Knowledge enhanced | – | Question difficulty |
| Bi-CLKT | 2022 | KBS | Joint CL, graph augmentation | Knowledge enhanced | – | Exercise-concept graph |
| HGKT | 2022 | SIGIR | Hierarchical graph neural network, RNN | Knowledge enhanced | – | Textual content + exercise graph |
| OKT | 2022 | EMNLP | GPT-2, ASTNN | Knowledge enhanced | – | Textual content |
| KET | 2023 | *Pattern Recognition Letters* | GNN | Knowledge enhanced | – | Exercise-concept graph |
| DKCT | 2023 | ADMA | Concept trees, multihead attention | Knowledge enhanced | – | Knowledge concept tree |
| S^2-HHN | 2023 | *Information Sciences* | Heterogeneous hierarchical graph, CL, RNN | Knowledge enhanced | – | Exercise-concept heterogeneous graph |
| RKT | 2020 | CIKM | Attention | Multiple | Forgetting behavior | Textual content |
| DGMN | 2022 | TKDE | Attention, GCN | Multiple | Forgetting behavior | Knowledge concept graph |
| QRAKT | 2023 | *Applied Sciences* | Attention | Multiple | Forgetting behavior + slipping and guess factor | Textual contnet + knowledge concept graph |
| KTMFF | 2023 | *Neural Computing and Applications* | Attention | Multiple | Speed | Textual content |

## Appendix E
Summary of Learning Analysis Methods

| Methods | Year | Venue | Techniques | Category |
|---|---|---|---|---|
| Kinnebrew et al. | 2013 | *Journal of Educational Data Mining* | Linear segmentation algorithm | Behavioral |
| Liu et al. | 2017 | *International Journal of Distance Education Technologies* | Sequential analysis | Behavioral |
| Han et al. | 2017 | ISET2017 | Lag sequential analysis | Behavioral |
| Liu et al. | 2017 | *International Journal of Distance Education Technologies* | Lag sequential analysis | Behavioral |
| Cheng et al. | 2018 | *Interactive Learning Environments* | Lag sequential analysis | Behavioral |
| Boroujeni & Dillenbourg | 2018 | Proceedings of the 8th International Conference on Learning Analytics and Knowledge | Clustering | Behavioral |
| Zhang et al. | 2018a | ICCSE | K-means, hierarchical clustering | Behavioral |
| Yan & Au | 2019 | *Asian Association of Open Universities Journal* | MLP, conjugate gradient | Behavioral |
| Wang et al. | 2019 | Proceedings of the 9th International Conference on Learning Analytics & Knowledge | Detailed access trajectories | Behavioral |
| Shen et al. | 2020a | Proceedings of the Seventh ACM Conference on Learning @ Scale | Bloom's taxonomy, social network analysis | Behavioral |
| Zhang et al. | 2020 | *Journal of Cloud Computing* | Deep belief network | Behavioral |
| DAISim | 2023 | CIKM | RL, GRU | Behavioral |
| Halawa et al. | 2014 | Proceedings of the Second European MOOC Stakeholder Summit | Logistic regression | Predictive |
| LadFG | 2016 | WSDM | Latent dynamic factor graph model | Predictive |
| Yan et al. | 2017 | IEEE Symposium on Visual Languages / Human-Centric Computing Languages and Environments | Machine learning classifiers | Predictive |
| Xing et al. | 2019 | *Journal of Educational Computing Research* | MLP | Predictive |
| E-LSTM | 2019 | ITME | LSTM | Predictive |
| RTP | 2019 | EDM | Recent temporal pattern | Predictive |
| CFIN | 2019 | AAAI | CNN, attention | Predictive |
| Goel & Goyal | 2020 | *Open Computer Science* | CNN | Predictive |
| Malysheva & Kelleher | 2020 | IEEE Symposium on Visual Languages/ Human-Centric Computing Languages and Environments | Random forest | Predictive |
| El Aouif et al. | 2021 | EIT | KNN, MLP | Predictive |
| DSLOP | 2021 | SDM | BERT, MLP, attention | Predictive |
| Worsley et al. | 2015 | ICMI | Affect- and pose-based segmentation | Both |
| L2S | 2017 | CIKM | Markov model | Both |
| Zhang et al. | 2017b | ISET | Logistic regression | Both |
| Sedrakyan et al. | 2020 | *Computers in Human Behavior* | Learning analytics dashboards | Both |
| Lu et al. | 2018 | *Journal of Educational Technology & Society* | Principal component regression | Both |